

\def\aa #1 #2/{{ A\&A}~{ #1}, #2}

\def\aarev #1 #2/{{ A\&A Rev}~{ #1}, #2}

\def\araa #1 #2/{{ ARA\&A}~{ #1},#2}

\def\aj #1 #2/{{ AJ}~{ #1}, #2}

\def\apj #1 #2/{{ ApJ}~{ #1}, #2}

\def\apjl #1 #2/{{ ApJ (Letters)}~{ #1}, L#2}

\def\apjs #1 #2/{{ApJS}~{ #1}, #2}

\def\mnras #1 #2/{{ MNRAS}~{ #1}, #2}

\def\qjras #1#2/{{ QJRAS}~{ #1}, #2}

\def\pasp #1 #2/{{ PASP}~{ #1}, #2}

\def\nat #1#2/{{ Nature}~{\ #1}, #2}

\def\physl #1 #2/{{ Phys.Lett.}~{ #1}, #2}

\def\physrep #1 #2/{{ Phys.Rep.}~{ #1}, #2}

\def\physrev #1 #2/{{Phys.Rev.}~{ #1}, #2}

\def\physrevb #1 #2/{{ Phys.Rev.B}~{ #1}, #2}

\def\physrevd #1 #2/{{ Phys.Rev.D}~{ #1}, #2}

\def\physrevl #1 #2/{{Phys.Rev.Lett.}~{ #1}, #2}

\def\sovastr #1 #2/{{ Sov.Astr.}~{ #1}, #2}

\def\sovastrl #1 #2/{{ Sov.Astr. (Lett.)}~{ #1}, L#2}

\def\commastr #1 #2/{{ Comm.Astr.}~{ #1}, #2}

\def\book #1 {{\it ``{#1}'',\ }}

\magnification=\magstep1

\def\frac#1/#2{\leavevmode\kern.1em \raise.5ex\hbox{\the\scriptfont0
#1}\kern-.1em /\kern-.15em\lower.25ex\hbox{\the\scriptfont0 #2}}

\def\lsim{\, \lower2truept\hbox{${<\atop\hbox{\raise4truept\hbox{$\sim$}}}$}\,}

\def\gsim{\,\lower2truept\hbox{${> \atop\hbox{\raise4truept\hbox{$\sim$}}}$}\,}

\def\ninit{\hoffset=.0 truecm \voffset=.0 truecm \hsize=15.5
	    truecm \vsize=23.5 truecm \baselineskip=14.pt %
	  \hsize=16. truecm \lineskip=0pt \lineskiplimit=0pt}



\ninit
\def\oneskip{\vskip\baselineskip}  

\centerline{\null}

\noindent
\centerline {\bf Sub--degree anisotropy observations:}
\smallskip
\centerline {\bf ground--based, balloon--borne and space experiments.}
\oneskip \parindent=25pt \parskip 0pt
\centerline{L. Danese$^1$, L. Toffolatti$^{2,5}$, A. Franceschini$^2$}
\smallskip
\centerline{M. Bersanelli$^3$ and N. Mandolesi$^4$}
\medskip\noindent
\llap{${}^1$}SISSA, Internat. School for Advanced Studies, 34014 Trieste, Italy
\medskip\noindent
\llap{${}^2$}Osservatorio Astronomico di Padova,
Vicolo dell'Osservatorio 5, I--35122 Padova, Italy
\medskip\noindent
\llap{${}^3$}IFCTR, CNR, via Bassini 15, I--20133 Milano, Italy
\medskip\noindent
\llap{${}^4$}ITESRE, CNR, via Gobetti 101, I-40129 Bologna, Italy
\medskip\noindent
\llap{${}^5$}Departamento de F\'\i sica, c.le Calvo Sotelo s/n, 33007 Oviedo,
Spain
\bigskip
\bigskip
\bigskip
\bigskip

\noindent
{\bf ABSTRACT} -- Extensive, accurate imaging of the Cosmic Background
Radiation
temperature anisotropy at sub--degree angular resolution
is widely recognized as one of the most crucial goals for cosmology
and astroparticle physics in the next decade. We review the
scientific case for such measurements in relation with sky coverage,
attainable sensitivity, confusing foreground radiation components, and
experimental strategies. Although ground--based and balloon--borne
experiments will provide valuable results, only
a well--designed, far--Earth orbit space mission covering a wide
spectral range and a significant part of all the sky will provide decisive
answers on the mechanism of structure formation.

\oneskip

\centerline{\bf 1. INTRODUCTION}
\medskip\noindent
A strong effort in observational cosmology will be dedicated in the next
years to measurements of Cosmic Background Radiation (CBR) anisotropies
on scales from several arcminutes to few degrees and on regions of the sky
as large as possible, to obtain the highly accurate statistics
required to answer fundamental questions about the origin of structures
in the universe. These observations would also have an enormous impact on
high--energy physics because of the linkage between studies
on primordial universe and those on particle physics.

As shown by many authors, the CBR anisotropy is
the result of many physical processes working on quite different scales
and cosmological epochs. Thus the  expected anisotropy is
sensitive to a number of initial conditions and physical
processes, such as the possible existence of field defects (walls,
cosmic strings, monopoles and textures), the slope and
amplitude of primordial energy density and gravitational wave fluctuations,
the Hubble constant, the cosmological constant, the baryon density,
the ionization history {\it etc.}
(Crittenden et al.\ 1993, Bond et al.\ 1993, Kamionkowsky et al. 1994).

Temperature fluctuations are usually decomposed in spherical harmonics
$$\Delta T/T\left(\theta ,\phi \right)\equiv \sum _{lm} a_{lm}
Y_{lm}\left (\theta ,\phi \right ),\eqno(1)$$
with rotational symmetric quantity
$a_l^2 \equiv \langle \mid a_{lm} \mid^2 \rangle$.
With this notation the all--sky variance is given by
$$\langle \left (\Delta T/T \right)^2 \rangle=\sum _l {2l+1\over 4\pi}a_l^{2}
W_l,\eqno(2)$$
where $W_l$ is a window function depending on the
characteristics of the experiment.

\smallskip
One of the most relevant open questions of the theory of structure formation
concerns the statistics -- gaussian or non--gaussian -- of primordial
anisotropies.

Density fluctuations generated by topological defects are in\-trin\-si\-cal\-ly
non--gaus\-sian and the mass distribution is rather sporadic. These
properties have been used in the attempt of explaining the strong
clustering of galaxies as compared to the mass distribution within an
Einstein--de Sitter universe.
Many authors pointed out that these models are severely
constrained by the $COBE$--DMR data, but still viable (Turok and Spergel 1990;
Bennet, Stebbins and Bouchet 1992; Ue--Li Pe and Spergel 1993).
The most effective way to discriminate between
non--gaussian and gaussian primordial perturbations is the
statistical analysis of the distributions of $\Delta T/T$
on angular scales ranging from several arcminutes to
one degree (Perivolaropoulos 1994; Coulson et al.\ 1994).

Cosmic strings are the most investigated case of topological defects and
many authors have produced detailed predictions. The
$COBE$--DMR data do not exclude string models (see e.g., Bennett, Stebbins
\& Bouchet, 1992). On the other hand, mapping the distribution of the CBR
temperature fluctuations in many thousands adjacent pixels
with angular resolution of $\sim 15^{\prime}$ would
enable a clear discrimination between gaussian and non--gaussian
primordial perturbations, if global uncertainties of at least $5\times 10^{-6}$
in $\Delta T/T$ per single pixel can be achieved (Hara et al.\ 1993).

Coulson et al.\ (1994) have produced 30$\times$30 degrees simulated maps
of CBR temperature fluctuations generated by cosmic defects.
All maps have been smoothed with a gaussian window of full-width
half--maximum (FWHM) $\sim 1^{\circ}$. It is apparent that
the differences among the case of strings, monopoles and textures are
only appreciable for experiments designed to
image the sky at high sensitivity and with large sky coverage.

\smallskip
Curvature (adiabatic) or isocurvature fluctuations might have been generated
at very early epochs.
Models based on the generation of gaussian, adiabatic fluctuations via
inflation have been extensively studied.
The standard inflation theory predicts that
the coefficients $a_{lm}$ and the rms fluctuations
are independent gaussian random variables. The theoretical predictions
yield estimates of $\langle a_{l}^{2} \rangle _{ens}$, the average
over an ensemble of universes (White et al.\ 1993).
As a consequence there is an inherent uncertainty in measuring
the radiation power spectrum $a_l^{2}$ even in the case of all--sky
surveys, since we are observing the fluctuation distribution
from only one point in a unique universe (cosmic variance).
For instance when we fix the normalization of the power
spectra by fitting the predicted $\Delta T/T$ to the COBE measurements we are
left with an uncertainty due to cosmic variance of $\approx$ 20$\%$
(White et al.\ 1993).

Small--scale
anisotropies are dominated by higher modes, which have more degrees
of freedom and smaller $\it cosmic$ uncertainty, because $a_{l}^{2}$
is $\chi ^2$-- distributed with $2l+1$ degrees of freedom (White, Krauss
and Silk, 1993). An additional source of uncertainty in reconstructing the
radiation power spectrum is connected to the sky coverage. Large sky coverage
at intermediate and small angular resolution
producing many thousands of pixels would squeeze the statistical errors
to few percent of the signal.
Scott et al.\ (1994) showed that the sample  variance
is approximately related to the cosmic variance by
$$\sigma_{sam}^2\simeq (4 \pi/A) \sigma_{cos}^2,\eqno(3)$$
where A is the solid angle covered by the observations.
Taking into account the dependence of the cosmic variance on the
pixel size (determined by the
beam angular resolution, $\theta_{beam}$) it can be shown
that the relative uncertainty due to the sample variance is
$\sim \left(2/N_{pix}\right)^{1/2}$,
where $N_{pix}\approx {A\over \theta_{beam}^2}$ is the
number of pixels in the observed sky area.
For instance, with $\sim$ 10,000 pixels available
(a conservative estimate for a well designed space
experiment) the relative uncertainty
in the coefficients $a_{l}^{2}$ is reduced to a few percent
at high confidence level (see Bond et al. 1993).
As for the expected level of the CBR fluctuations on half--degree angular
scale, both theoretical models and available observations
predict values of $\Delta T/T\sim $few$\times 10^{-5}$
(see e.g. Smoot et al.\ 1992; Gundersen et al.\ 1993; Cheng et al.\ 1994; Bond
et al.\ 1993; Sugiyama et al. \ 1993).

Imaging the CBR fluctuations  with high precision and
accuracy over a {\it large} portion ($\sim$ 1/4--1/3) of the sky
at angular resolutions ranging
from {\it 10' to 30'} will not only discriminate between gaussian
and non--gaussian initial perturbations, but will also produce
fundamental observations to investigate
the history of the universe.

Recently Bond et al. (1993) examined to what extent
the initial conditions and the evolution of the
perturbations in the universe can be inferred from the analysis of
CBR anisotropy. They showed that the determination of the CBR anisotropy
on angular scales from about 10' to few degrees would afford important
information on a combination of basic parameters such as the power
law index $n_s$ of the amplitude of the
energy density fluctuations, the ratio $r$ of the tensor--to--scalar
quadrupole,
the Hubble constant, the cosmological constant and the epoch
of the reionization. The reconstruction of the radiation power spectrum
together with independent observations on large--scale structures,
on the Hubble and cosmological constants and the thermal history
of the universe, would allow to test the inflation hypothesis.

It is expected that in the next decade degree and sub--degree anisotropy
measurements will be performed with a variety of techniques and
experimental designs. The main focus of this paper is to compare the
relative merits and challenges of experiments designed
to be performed from ground, balloon and space.
In \S 2 we will discuss the problems posed by foreground
emissions. In \S 3 we briefly address the main experimental approaches.
After a discussion in \S 4 we present our conclusions in \S 5.


\oneskip \centerline{\bf 2. FOREGROUND EMISSIONS}
\oneskip\noindent
The foreground
radiations hampering the measurements of the CBR anisotropy are
the galactic emission and the integrated emission from extragalactic
unresolved sources. Atmospheric
emission is an additional foreground source for ground--based and
balloon--borne observations. In order to greatly reduce the
uncertainties in observing CBR anisotropy one has to accurately subtract
the contributions by unwanted foreground radiations.

Instrumental sensitivities at few $\mu$K level with reasonable
integration time are now reachable. As for the galactic contribution,
the limiting factor in interpreting the
measurements is our knowledge of galactic emission even
in the case of observations with intrinsic sensitivities at the level of
several tens of $\mu$K (the $COBE$--DMR sensitivity; see Bennett et al.\ 1992;
1994).

\oneskip\centerline {\bf 2.1  Atmospheric emission}
\oneskip \noindent
Atmospheric emission is the largest unwanted emission
for ground--based and balloon--borne experiments.

For 40 km altitude  balloon flights the expected atmospheric continuum
emission at the zenith yields an antenna temperature of a few mK.
On the other hand, with the large bandwidths used by bolometric techniques
the line emission can give an important additional contribution to the
atmospheric signal. As a result, the atmospheric subtraction is not so trivial.
For instance for the TopHat experiment the chopping technique has been
designed for monitoring the time--variable atmospheric noise (Cheng 1994).

For ground--based experiments much depends on frequencies and locations.
To be specific let us refer to the Tenerife site and to the
Amundsen--Scott South Pole station.

At Tenerife (Canary Islands)
will be located the Very Small Array (Lasenby et al.\ 1994)
experiment, an interferometer
that will map the CBR.
The site is 2.4 km above sea level. At this altitude and with
water vapor column of 2 mm (typical value for that site)
the antenna temperature of the atmosphere  is $T_A\sim 4.7
\ K$ and $T_A\sim 8.7\ K$ at $\nu=28$ GHz and
$\nu=38$ GHz respectively. These antenna temperatures are
rather similar to the antenna temperature expected at Cambridge
at 15 GHz, the operative frequency of the
prototype Cambridge Anisotropy Telescope
(CAT). CAT observations seem to be unaffected by atmospheric noise down
to $\sim 2\times 10^{-5}$ (Efstathiou, private communication; Lasenby
et al.\ 1994), demonstrating that the
atmosphere on arcmin scales does not produce noise at levels larger
than $4\times 10^{-6}$ $T_A$. Even assuming than on larger scales the
atmospheric fluctuations do not increase, we would need
to lower the effects of atmospheric noise at least  by a factor
$\sim 10$ to get the precision required to disantangle the
contribution of the galactic foregrounds. The possibility
of obtaining such a level of subtraction of the atmospheric
emission is still unsettled. For instance, in the case of the ACME experiment
at South Pole (possibly the best site in terms of atmospheric emission and
stability) at frequencies around 30 GHz, the
main limitation is due to the atmospheric noise (Gaier et al.\ 1992).
After removing $\sim 70$\% of the data for {\it bad} weather a limit of
$\Delta T/T\leq 1.4\times 10^{-5}$ was found. Since the atmospheric emission
amounts to about 4.6 $K$, with only 0.15 $K$ contributed by water vapor
(0.5 mm $H_2 O$ column), the atmospheric noise should be at a level of
$3\times 10^{-6}$ $T_A$.

In addition to water vapor fluctuations, pressure gradients
in the observed sky patches are likely to induce significant variations
of the O$_2$ emission, as direct measurements from the South Pole
site have shown (Meinhold \& Lubin, 1991; Meinhold et al 1993;
Bersanelli et al 1994).

In conclusion it may be a very hard test for ground--based experiments
to remove atmospheric noise to the levels required to accurately
subtract the galactic foregrounds on large portions of the sky.
Even for balloon--borne experiments this issue proves not trivial.

\oneskip\centerline{\bf 2.2  Galactic emissions}
\oneskip \noindent
The galactic radiation at frequencies below $\sim 20$ GHz is
dominated by the synchrotron emission which approximately follows
a power law in frequency
$$ I_{sync}\propto \nu ^{-\beta}\,\eqno(4)$$
with $\beta\sim 0.7-0.9$ and a possible steepening towards higher frequencies
(see Bennett et al.\ 1992). All--sky maps at low frequencies are
available at 408 and 1420 MHz (Haslam et al.\ 1970; Haslam et al.\ 1974;
Haslam et al.\ 1982; Reich 1982; Reich \& Reich 1986, 1988).
An extensive map at 19 GHz has been obtained by Boughn et al. (1992).
Complementary studies on
polarization, magnetic fields and electron energy distributions have been
performed. As a result the contribution of the synchrotron emission
is rather well understood, although much work still has to be done.
A significant improvement is expected in the near future as measurements
currently in progress will yield multifrequency galactic maps
over lage sky areas (e.g. De Amici et al. 1994).

\noindent
Our knowledge of the free--free emission, the other galactic foreground at
radio wavelengths, is at the moment rather poor. However, the free-free
emission seems to be concentrated towards the
galactic plane (Bennett et al., 1992), so that its importance is
partially reduced at high glactic latitudes.
It would significantly dominate on the synchrotron only at frequencies
$\nu \gsim 100$ GHz due to its
frequency dependence ($I_{ff}\propto \nu ^{-\alpha}$, with $\alpha \sim 0.1$).
However, at these frequencies
the galactic dust emission already dominates.
As a consequence no direct maps of the free--free emission are available
and only indirect reconstructions have been produced
by subtracting the synchrotron component from radio maps
(Bennett et al 1992; 1994).
Additional indirect indications on free-free emission come from the
relationship of the free--free emission with other probes
of the warm ionized medium
such as the $H_{\alpha}$ and $N^{+}$ emissions (Reynolds 1992; Wright et al.\
1991; Bennett et al.\ 1992; Bennett et al.\ 1993).

Dust dominates the galactic emission at frequencies $\nu \gsim$ 90--100 GHz.
Dust emission depends on the Interstellar Radiation Field, gas chemical
composition, dust to gas ratio, and grain composition, dimension and structure.
Thus variations from place to place are expected.
Relevant information comes from the IRAS 60 and 100 $\mu$m surveys
with few arcminutes of resolution, from the
$COBE$--FIRAS experiment with resolution of about 7 degrees (Wright et al.\
1991), from the map at 170 GHz with resolution of 3.8 degrees
produced by the MIT survey (Ganga et al.\ 1993; Meyer et al.\ 1991).

About dust emission spectrum and fluctuations additional information
is expected from the analysis of the far--IR $COBE$--DIRBE maps
(Hauser 1993).

Wright et al.\ (1991) presented their results in terms of a product
of a function of position times a function of frequency:
$$I_G(l,b,\nu)=G(l,b)g(\nu).\eqno(5)$$
The best fit of the frequency function turned out to be:
$$ g(\nu)\approx 2.2\times 10^{-4} \left (\nu/900 GHz\right)^2
\times \left[B_{\nu}(20.4)+6.7B_{\nu}(4.77)\right].\eqno(6)$$
where $B_{\nu}$ represents the Planck function.
As for dust anisotropies, Gautier et al.\ (1992) studied the angular power
spectrum of the IRAS 100$\mu$m maps and concluded that
dust fluctuations decrease with angular scale according to a power low
with an index $\sim 0.45$.



\oneskip\centerline {\bf 2.2.1 Fluctuations of the galactic emission at
sub--degree angular scales}
\oneskip \noindent
The level of the expected galactic noise can be predicted from the already
available data. Brandt et al.\ (1993), starting from the radio maps at
408 and 1420 MHz and using all other available information (e.g. spectral
indexes, magnetic fields etc.), were able to pick out 50
high galactic latitude regions of
10$\times$10 degrees in which synchrotron and dust emission at 31 and 100 GHz
are rather smooth on angular scales of $\sim 1$ degree.
Actually, in these regions the rms intensity variations are less than
$2.1\times 10^{-4}$ MJy/sr with a total intensity ranging
from 6 to 10 times the rms. Figure 1 shows that for these regions
the expected synchrotron rms variations (dotted line) yield
$\left(\Delta T/T\right)_{rms}\lsim 3\times 10^{-6}$ at
frequencies $\nu \gsim 40$ GHz.
Therefore, in the frequency range $50\lsim \nu \lsim 600$ GHz,
the synchrotron intensity fluctuations
are significantly smaller than CBR fluctuations.

Similar conclusions can be drawn for the case of the free--free
emission. Using the free--free to $H_{\alpha}$ and $N^{+}$ emission ratio
(Wright et al.\ 1991; Reynolds 1992; Bennett et al.\ 1992) it is
possible to infer a generous estimate of
$1.2\times 10^{-3}$ MJy/sr at $\nu=53$ GHz for the total free--free emission
at $\mid b\mid\geq 40$ (which corresponds to $\sim 15,000$ square degrees).
{}From Figure 2 of Reynolds (1992) rms variations of about 30$\%$ of the
total emission can be estimated. Figure 1 shows that the corresponding
$\Delta T/T$ level (short--dashed line) adds a small contribution to the
expected primeval CBR fluctuations at frequencies $50\lsim \nu\lsim 500$ GHz.

At still higher frequencies, due to rapid decrease of the CBR intensity,
the radio emissions would
provide again detectable anisotropies but they are completely swamped
by the anisotropies produced by dust emissions (see Figure 1).

The $COBE$--FIRAS dust map (Wright et al.\ 1991) shows that a non negligible
portion of the sky exhibits rather low brightness at sub--mm frequencies.
Taking the brightness
at 100 $\mu$m as a reference, 10$\%$ of the sky shows a brightness
less than 1.5 MJy/sr. Dust emission can be indirectly derived from HI maps
(see e.g. Stark et al. 1992). As reported by  Lockman et al.\
(1986) about 8$\%$ of the sky has HI column densities $N_H\lsim 1.5
\times 10^{20}\ cm^{-2}$. Using the correlation between 100 $\mu$m and
HI emission derived by Boulanger \& Perault (1988) such column densities would
correspond to sky brightness less than 1.3 MJy/sr at 100 $\mu$m.
The distribution of the
column densities is rapidly increasing from $N_H\sim 5\times 10^{19}$,
which is the lowest detected value in the sky. No more than 1$\%$ of the
sky exhibits $N_H\leq 1\times 10^{20}$. Extremely--low emission regions,
such as the Lockman Hole or the South Pole Hole
or the GUM area, are rare (Lockman et al.\ 1986).

It is worth noticing that areas of the sky with relatively low HI column
densities are also remarkably smooth, with a dispersion
$\sigma/\langle N_H\rangle \lsim 0.15$ (Lockman et al.\ 1986).

In Figure 1 we reported the fluctuations generated by the galactic dust
based on the assumption that the rms variations amount to 100$\%$  of
the total emission in the lowest emission regions,
i.e. 0.4 MJy/sr at 100 $\mu$m (see above), a level
which is well within the capabilities of a high sensitivity survey for
dust mapping in the sub--mm range. 
It is apparent that the predicted dust fluctuations are
rather small $\left(\Delta T/T\right)_{rms}\lsim 5\times 10^{-6}$ for
frequencies $\nu \lsim 300$ GHz.

At high enough galactic
latitude ($\mid b\mid\gsim 40$) subtraction of the galactic
emission is on average required with  relative accuracy levels
of $\sim 20\%$ of the signal
in order to get galactic noise below several $\mu K$
in the frequency range $30\lsim \nu \lsim 300$ GHz.
To reach  this goal large
sky coverage, a few $\mu K$ sensitivity and wide frequency range
are needed.


\oneskip\centerline {\bf 2.3  Extragalactic emissions}
\oneskip \noindent
A short but complete report on source confusion due to extragalactic
point sources can be found in Toffolatti et al. (1994).
Their results show that
extragalactic point sources are expected to produce
$\left(\Delta T/T\right)_{rms}\lsim 2\times 10^{-6}$ at angular scales
of interest here ($10^{\prime}\lsim \theta \lsim 30^{\prime}$)
in the fre\-quen\-cy ran\-ge $50\lsim \nu\lsim 300$ GHz
(see Figure 2).
As previously discussed by Franceschini et al. (1989), the main
source of uncertainty in predicting the confusion noise due to extragalactic
sources is the very large frequency extrapolation
of the source counts, which are currently known only at cm wavelengths
in the radio and at 60 $\mu$m in the far--IR.

Referring to Franceschini et al. (1989) we can say that the present
predictions should be accurate within $\approx 30$\% for $\lambda \gsim 1$
cm and to within a factor of $\sim$2 down to $\lambda\approx$ a few
millimeters. In the sub--mm wavelength range, where the confusion noise
is dominated by dust emission, the two adopted models give an appropriate
estimate of the uncertainty in the predictions (see caption to Figure 2).
The limits reported in Figure 2 can be achieved
eliminating 5$\sigma$ sources (see Franceschini et al.,
1989 for more details). If source surveys were available with
limiting fluxes fainter than 5$\sigma$, then the source contribution can be
lowered and the optimum frequency range widens.
Fluctuations generated by galaxy clusters through the S--Z effect
provide foreground noise in experiments designed to map the
intrinsic CBR anisotropy. Recently
Bond \& Myers (1993) have investigated the
formation and evolution of galaxy clusters in the frame of CDM
models. Indeed they considered not only the {\it standard}
CDM model, but also other alternatives of structure
formation within the inflationary scheme. They
have produced the angular power spectra of the S--Z effect
produced by various structure formation models. Their
results show that the integrated S--Z effect would
produce fluctuations at level of 10$\%$ of the primary
$\Delta T/T$ at scales $> 10^{\prime}$. Thus
S--Z effect would not affect the detection of primordial fluctuations
on larger scales.
Ceballos \& Barcons (1994) 
using a parameterization of the X--ray luminosity function of clusters
and a simple model for its evolution were able to show that the average
Compton $<y>$ parameter is always $< 10^{-6}$. Similar, albeit a little
higher, values are also found by Markevitch et al. (1992),
who combined X--ray measurements of the luminosity distributions of
relatively nearby clusters ($z< 0.2$) with simple models of structure
formation and cluster evolution.

Ceballos \& Barcons (1994) also show that the
anisotropies imprinted at arcmin scales are dominated by the hottest
undetected objects and that they are negligible ($(\Delta T/T) \lsim 10^{-6}$)
at $\lambda \gsim 1$ mm while becoming more important at shorter wavelengths
($(\Delta T/T) \sim 10^{-5}$ at $\lambda \simeq 0.3$ mm).
Anyway, since most clusters
will produce an isolated and detectable feature in the sky maps they could be
subtracted out leaving only less bright objects producing negligible
background noise. For beamsizes $\sim 10$ arcmin confusion will be more
relevant and only the hottest clusters ($T_{gas}\gsim 7$ keV) will be
detectable. Even in that case, the remaining undetected objects will
produce negligible sky noise.
On the other hand, Bond \& Myers (1993) pointed out that
the S--Z effect generates a significant non--gaussian tail
in  $\Delta T/T$ distributions on scales of about 10$^{\prime}$
that could be probed provided that large enough statistics are available.
A coverage of several $10^4$ pixels is required to get significant results
(see Bond \& Myers, 1993 for more details).

Due to the nature of the S--Z effect
the deflections are negative for frequencies $\nu \lsim 200$ GHz
and positive at higher frequencies.
Thus experiments in the appropriate frequency range  might test the
non-gaussianity generated by S--Z effect.

\oneskip\centerline{\bf 3. REMARKS ON EXPERIMENTAL APPROACHES}
\oneskip\noindent
The different environment conditions which characterize ground-based,
balloon-borne and space experiments strongly affect the choice of the measuring
technique to be implemented, and determine the
limitations and quality of the expected results. Table I summarizes the
main aspects discussed here.

\oneskip\centerline{\bf 3.1  Frequencies and techniques}
\oneskip\noindent
As mentioned, ground based measurements are allowed only in few fixed
atmospheric windows downwards of 90 GHz, and only few independent
frequency bands can be covered. In this regime one can take advantage of
the new generation, high-quality, coherent detectors, such as SIS
(Superconductor--Insulator--Superconductor) (see Pan et al., 1989) or, even
better, HEMT (High Electron Mobility Transistors) radiometers (Pospieszalski,
1993; Pospieszalski et al., 1993). These devices have very low noise figures,
especially when cryogenically cooled, and have proved suitable for these
experiments (see e.g. Meinhold et al., 1993).

In the case of balloon experiments the effects of the
atmosphere are strongly reduced (although not suppressed) so that
high--frequency measurements ($100<\nu < 900$ GHz) can be performed.
Very sensitive cryogenically cooled
bolometers are the typical solution for balloon--borne experiments (see Figure
3). The high sensitivity is achieved with broad--band channels which,
however, decrease the power in the spectral separation
(typically the 3-4 band allowed are adjacent to each other).

For a space mission, in principle, the whole frequency range is available
with no limitations other than the unavoidable galactic and extragalactic
emission (see \S 2). The European Space Agency COBRAS/SAMBA project
(ESA M3 Assessment Study Results, Summary Reports SCI(94)9, May 1994) takes
full
advantage of this possibility proposing measurements in the frequency range
$30<\nu <900$ GHz by integrating radio-- and bolometric detectors at the focal
plane of a $\sim 1.5$ m aperture off--axis telescope on a payload in a
far--Earth orbit.
Passively cooled receivers with ultra--low noise HEMT
preamplifiers can cover the low frequency interval up to $\sim 130$ GHz,
while bolometers cooled down to $\sim$ 0.1 K
provide extremely sensitive detectors in the range 100--900 GHz.
The two different detection techniques can easily
exploit 7-8 observing bands, with the possibility of frequency overlap close
to 120 GHz.
This, together with large sky coverage, allows strong control over
systematic errors by a very powerful frequency/foreground/technique
crosscheck of the maps.

\oneskip\centerline{\bf 3.2 Contamination from Earth radiation}
\oneskip\noindent
Sub--degree anisotropy experiments carried out from the ground or
near the Earth's surface require very efficient rejection of
off-axis radiation to avoid comtamination from the
unwanted emission from the Earth. The level of rejection required
becomes increasingly severe when the goal sensitivity
is pushed to more and more ambitious limits. From balloon altitudes,
since the distance of the  gondola from the Earth
is negligible compared to the Earth radius, the sidelobe and straylight
rejection required is of the same order as for ground-based experiments.

If one imposes the total ground contribution
to fall below significance level,
a ground--based or balloon experiment with angular resolution $\sim 30'$
and sensitivity $\Delta T/T \sim 10^{-6}$
implyes a rejection factor at large angles as high as 10$^{12}$
to 10$^{13}$.
However, since the measurement is contaminated by
{\it variations} of the ground radiation spill--over rather than its
{\it absolute value},
this factor should be regarded as a conservative limit
for a worse--case scenario. In principle, significantly less
stringent rejection limits (probably by a factor $\sim 100$)
can be acceptable provided that the experiment does not require
movements, relative to the Earth, of the instrument parts
which control the amount of ground radiation adding to the measured
signal. Typically, this is the case of ground--based experiments, which
also have the advantage of possible use of extended reflective
gound screens to minimise the signal.

While meaningful measurements from the ground or from balloons have
been performed with accuracies $\Delta T/T \sim 10^{-5}$, Earth radiation
contamination has always been a major concern. Pushing the sensitivity
to the next order of magnitude is likely to be very problematic,
particularly with the objective of extended sky coverage.

A space mission from a low--Earth orbit would only marginally relax this
critical requirement, since the Earth would still
cover about $1/4$ of the total solid angle. For instance,
microwave emission from the Earth has
been a serious concern in the design and
systematic error analysis of the $COBE$--DMR experiment,
even at the relatively broad angular
resolution ($7^{\circ}$) of its antennas
(Kogut et al., 1992). From the $COBE$ 900 km circular orbit, the Earth is
a circular source with angular diameter
122$^{\circ}$ and minimum temperature
285 K. Upper limits to the antenna temperature of
the Earth signal contribution
to the DMR 2--years data are at level 25--60 $\mu$K (95\% confidence
level), (Bennett et al., 1994).
It is clear that rejection of Earth radiation
is a challenge to low--Earth orbit experiments aiming to reach
sensitivities a factor $\sim$ 10 better than $COBE$--DMR with
beam areas smaller by a factor of $\sim$100 to $\sim$1000.

Only by moving the payload to a far--Earth orbit, like Moon--Earth libration
point L5 ($\sim 400,000$~km, i.e. the orbit presently selected
for COBRAS/SAMBA), or Sun--Earth libration point L2
($\sim 1,500,000$~km, i.e. the orbit proposed for the PSI
mission, and also considered for COBRAS/SAMBA),
the Earth's solid angle is
greatly reduced, thus decreasing by the same factor the required rejection.
In the most conservative approach, the rejection requirement
drops by four order of magnitude ($10^{8}$ to $10^{9}$), and
becomes comparable to that needed
to suppress Sun radiation. This rejection level can be obtained
with careful, though conventional design of the optics and shielding.

An additional advantage of a deep--space environment concerns the
thermal conditions ($T\sim 100$~K, stability $\simeq$ 0.1 K/few weeks)
which are ideal for passively cooled,
 long-lifetime experiments. For example, the COBRAS/SAMBA
Low Frequency Instrument (LFI) is a multifrequency radiometer array designed to
function for up to four years exploiting the passive cooling technique,
although the scientific objective will be reached within the nominal 2
years mission lifetime.

\oneskip \centerline{\bf 3.3 Experimental configurations}
\oneskip\noindent
The main design considerations are driven by
the need to obtain the nominal experiment performances
(resolution, sensitivity, sky coverage, window function, etc.),
with minimum and well--controlled potential systematic effects.

In order to achieve angular resolution $<1\deg$ in the
frequency range allowed by galactic and extragalactic foregrounds
($30\lsim\nu\lsim 300$ GHz)
large ($\sim 1$~m) telescopes are needed, which properly
redirect in the sky the feed beam pattern.
Clear aperture, oversized optics, either Gregorian or Cassegrain,
are used to minimize diffraction and spillover effects
(e.g. Meinhold et al., 1993; Fisher et al., 1992).

In most sub--degree anisotropy experiments, a moving (rotating or nutating)
subreflector provides the beam--switching pattern
necessary to efficiently subtract instrumental and
atmospheric drifts. The Tenerife triple--beam switch strategy, suitable
for their larger ($\sim 8\deg$) angular scale, has proved particularly
effective for the subtraction of slow atmospheric drifts (Davies et al. 1992).
In general, moving the subreflector or other reflective parts is a source
of concern for potential systematic effects, such as
possible modulation of signals from the Earth (or the balloon)
diffracted in the beam or sidelobes by the instrument structure.
In general, the rejection and control of systematic errors imposed by
a sub--degree measurement at sensitivity $\Delta T/T \approx 10^{-6}$
requires an environment and viewing field extremely
free from local contamination. An eloquent example of the
critical level reached by traditional balloon or ground
based designs for these experiments is the futuristic concept
proposed for TopHat (Cheng 1994).

The window function of the experiment, which is determined by
the instrument beam and sky scanning technique, selects
the angular range at which the measurements are most sensitive.
Typical beam switch experiments have window functions $W_l$
peaking in a relatively narrow range
of the corresponding scales of the primordial power spectrum
(Bond et al. 1991). Most performed measurements have
been designed to sparsely sample the autocorrelation function at
a fixed angular scale; on the other hand, to
obtain a model--free reconstruction of the power spectrum
one needs a filter function $W_l$ nearly constant
over a range of angular scales as wide as possible.
This is the characteristic of an imaging observation
(Readhead and Lawrence 1992), where
several widely different angular scales are simultaneously probed.
Interferometry and arrays of radiometers or bolometers,
with proper sky scanning strategy,
can both in principle obtain images of large areas of the sky
with the required sensitivity.

\oneskip\oneskip \centerline{\bf 4. DISCUSSION}
\oneskip \noindent
Large sky coverage and high final sensitivity are
fundamental requirements for measurements
of sub--degree CBR anisotropy aiming to cast light on the
nature and the origin of the fluctuations from which the present
stuctures in the universe have developed.
Several lines of argument point to quantify
the sky coverage requirement to at least 10,000--15,000 square
degrees or, equivalently, about 30$\%$ of the sky and
the sensitivity to $\Delta T/T \approx 10^{-6}$.
For the case of an imaging experiment, the large number of pixels would
permit to precisely reconstruct the radiation power spectrum from the
beam angular resolution up to the largest scale allowed by the sky sampling.
In the case of the COBRAS/SAMBA experiment this will be obtained in the
angular range from $10^{\prime}$ up to the $COBE$--DMR angular resolution.
Large and high quality statistics would
also make possible to discuss the contribution of the galaxy clusters
to the fluctuations distribution, with the aim of discriminating
among various possible histories of their evolution.


Are these goals reachable with ground--based or balloon--borne experiments?

As for ground--based experiments three major difficulties have to be
solved that are somewhat connected: atmospheric noise, galactic emission
removal and sky coverage.


Since above 25 GHz the atmospheric signal at the best ground--based sites
(e.g. at the Amundsen--Scott South Pole Station) ranges  from
$\sim 3.5$ to 9 $K$, we need to eliminate the atmospheric noise
to a level $\lsim 10^{-6}$ of the signal. This is a factor of 3--4
smaller than the performances of current experiments which already run
into the atmospheric limit (Gaier et al.\ 1992).
In addition to the well known water vapor
variations, high and low pressure regions
in the atmosphere are likely to determine significant
O$_2$ fluctuations which require
very long integration to be averaged out (Meinhold et al. 1993).

The Galaxy emission removal is strictly related to the problem of the
atmospheric noise and observing frequency.
In general  precise subtraction of the galactic contribution
is very difficult with observations confined
in a small frequency range and in the presence of atmospheric noise
(Brandt et al.\ 1993).
{}From Figure 1 it is clear that even in the regions of the sky in which
the emission is low, subtraction of the galactic signal
is needed at frequencies $\nu \lsim 40$.
Unfortunately, the possible observing frequencies
from the ground are restricted by the presence of
the strong H$_2$O line at 22.2 GHz and the oxygen
band peaking at about 60 GHz (see e.g. Danese \& Partridge 1989).
Thus atmospheric noise and the limited range
of available frequencies conspire against precise
removal of the Galaxy contribution. In order to minimize galactic foreground
emissions, measurements have been attempted in the atmospheric window near
$\nu \simeq 90$~GHz. However at this frequency the atmospheric antenna
temperature is  $\gsim 9$~K even at the Amundsen--Scott South Pole Station.

Attempts of obtaining large sky coverage from the ground face additional
problems. Because both the total contribution and
atmospheric fluctuations strongly increase away from the zenith,
ground--based observations are confined in regions around the
zenith. This constraint also limits the possibility of chosing
low--foreground sky patches for the observations.

While it is conceivable that detection of anisotropies
with amplitude $\Delta T/T \sim$ 3--4$\times 10^{-5}$
can be obtained from the ground at 3--4 $\sigma$ level,
nonetheless it is clear that the large sky coverage
required by the arguments presented in \S 1 is not
obtainable.

\smallskip
As for balloon--borne experiments the main problem is the sky coverage,
which is strongly limited by the short flight time.
Even using very sensitive bolometers such as those foreseen
for the  upgraded MAX experiment,
it seems possible to get only about 50--100 pixels per flight.
TopHat and similar experiments are presently under study,
with the aim of minimizing
the instrumental systematics. Polar long--duration flights are also
foreseen. On the other hand assuming very optimistically
that these experiments
successfully flew 4--5 times in the next 10 years, they
would produce data for no more than 500--1000 pixels (see, e.g., Cheng
1994). As a consequence the large coverage needed to answer the fundamental
cosmological questions are well beyond the capabilities of balloon--borne
experiments flying in the next 10 years.

In smaller sky patches high-quality balloon measurements are
possible, most likely at high ($>90$~GHz) frequencies.
However, as pointed out in \S 3, balloon measurements
at sensitivities $\Delta T/T \sim 10^{-6}$
will have to suppress earth radiation
with an efficiency 3 to 4 order of magnitude more than what
can be presently achieved. Moreover, the residual atmospheric contamination
and the small coverage enhance the problem of the Galaxy emission removal.
As mentioned before, the Galaxy subtraction problem is
alleviated only when observing the {\it few low emission} regions of the sky.
On the other hand, the access to these regions from balloons
during long--duration flights might be difficult.

A further point to take into account is that there is no substitute
for {\it imaging} observations (Readhead \& Lawrence 1992).
Indeed only by imaging the CBR anisotropies
the power spectrum can be analysed in a model--independent way.
While single or double switching procedures
used in balloon--borne experiments can only sparsely sample
the power spectrum, imaging free from systematic errors
is obtainable only with a space mission.

\smallskip
In conclusion accurate studies of the CBR anisotropy power spectrum
from $10^{\prime}$ to degree angular scales
are possible with the data obtainable only with
a space mission in a far-Earth orbit.
In particular the joint COBRAS/SAMBA mission has many exciting
characteristics:
\item{$\bullet$} it would cover {\it all} the frequency range
($50<\nu<300$ GHz) and {\it all} the
sky regions (basically $b\gsim \mid 30\mid$ i.e. $\sim$ 1/2 of the entire sky)
for which the subtraction of the foregrounds would not
give troubles for
high sensitivity all sky surveys from space; this would results in negligible
statistical errors in the power spectrum reconstruction up to a few degrees;
\item{$\bullet$} it will be placed in a far--Earth orbit ($\sim$ 400,000 km
from the Earth), dramatically decreasing the problem of the sidelobe and
straylight rejection;
\item{$\bullet$} at low frequencies (observations performed by radiometers)
the galactic noise is generated by synchrotron and free--free emission,
while at the highest frequencies of the experiment (observations performed
by bolometers) dust emission dominates.
Indeed this mission would get 5--6 maps of the CBR fluctuations
with high angular resolution on more than half of
the sky and in the best frequency window, where galactic and extragalactic
emissions are at their minimum value,
as it is apparent from Figure 1 and 2. This would offer the opportunity of a
fundamental cross-check of the results;
\item{$\bullet$} at three frequencies (31, 53 and 90 GHz) the product maps
can be straightforwardly compared with those of $COBE$--DMR;
\item{$\bullet$} although confined to the tails of the fluctuation
distributions, nevertheless the S--Z effect in galaxy
clusters produce different features
(positive and negative tails respectively) on the fluctuations distributions
at  different
COBRAS/SAMBA frequencies, allowing a clear separation of the effect
from other possible sources of non--gaussianity. In this respect it
will be extremely interesting to compare the COBRAS/SAMBA maps
to the ROSAT all--sky survey of galaxy clusters.


\oneskip\oneskip \centerline{\bf 5. CONCLUSIONS}
\oneskip \noindent
A wealth of information
about the history of the Universe from the very beginning to the
present structure is  imprinted in the CBR anisotropy.
The observations and the theoretical work
so far done have spectacularly increased our knowledge in the
field and have stimulated a number of observational projects for
the next decade. The proposed experiments will be carried out from
ground, balloon and space.

The above discussion has shown that only a precise and accurate
reconstruction of the CBR power spectrum on angular scales
ranging from several arcminutes to
several degrees will produce an exceptional and possibly decisive
step towards the understanding the physical state of the very early
universe and the evolution to its present structure,
with enormous impact on particle physics.
Large sky coverage and precise foreground subtraction (which indeed
are deeply interrelated) are mandatory to achieve this goal.
Only by imaging CBR anisotropies over a large fraction of the sky we will
obtain a model-independent, precise determination of the
radiation power spectrum.

Although it is apparent that much can be learnt
from ground--based and balloon--borne experiments, nevertheless
they are not suitable to produce extended maps of the sky to the required
precision in the next ten years or so.
Only from space a properly designed experiment can image the sky
with the needed precision and coverage.
The frequency range, sensitivity and orbit proposed by the COBRAS/SAMBA
project will allow to produce near--all--sky
maps with very small systematic errors, which
will be one of the most fundamental data set in cosmology
and astroparticle physics.


\bigskip\bigskip

{\bf Acknowledgements} L.T. would like to thank the Department of Mathematics
of the University of Oviedo (Spain)
for the hospitality during part of the preparation of this paper.
This work has been partially supported by the Commission of the European
Communities, ``Human Capital and Mobility Programme'' of the EC, contract
number CHRX--CT92--0033 and by the Agenzia Spaziale Italiana (ASI).

\vfill\eject
\oneskip \noindent

\centerline{\bf REFERENCES}

\oneskip
\hyphenpenalty=5000

\def\ref#1\par{\noindent \parshape=2 0in 12.65cm 1cm 11.65cm{#1} \par}

\ref
Bennett, C.L., et al., 1992, 396, L7

\ref
Bennett, C.L., Stebbins, A., \& Bouchet, F.R., 1992, ApJ, 399, L5

\ref
Bennett, C.L., Hinshaw, G., Banday, A., Kogut, A., Wright, E.L.,
Loewenstein, K., \& Cheng, E.S. 1993, ApJ, 414, L77

\ref
Bennett, C.L. et al.\ , 1994, COBE preprint

\ref
Bersanelli et al. 1994, ApJ, in preparation.

\ref
Bond, D., et al. 1991, Phys. Rev. Lett. 66, 2179.

\ref
Bond, J.R., Crittenden, R., Davis, R.L., Efstathiou, G., \& Steinhardt,
P.J., 1994, Phys. Rev. Lett., 72, 13

\ref
Bond, J.R., \& Myers, S.T., 1993, preprint.

\ref
Boughn, S.P., Cheng, E.S., Cottingham, D.A., \& Fixsen, D.J., 1992, ApJ,
391, L49

\ref
Boulanger, F., \& Perault, M., 1988, ApJ, 330, 964

\ref
Brandt, W.N., Lawrence, C.N., Readhead, A.C.S., Pakianathan, J.N., \&
Fiola, T.M. 1993, ApJ, in press

\ref
Ceballos, M.T., \& Barcons, X. 1994, MNRAS, in press

\ref
Cheng, E.S., Cottingham, D.A., Fixsen, D.J., Inman, C.A., Kowitt, M.S.,
Meyer, S.S., Page, L.A., Puchalla, J.L. and Silverberg, R.F., 1994,
ApJ, 422, L37

\ref
Cheng, E.S., 1994, Proc. of the Workshop "Present and Future of
the Cosmic Microwave Background", Santander, Spain, 1993, eds. J.L. Sanz,
E. Martinez--Gonzalez \& L. Cayon, Springer--Verlag, p.76.

\ref
Coulson, D., Ferreira, P., Graham, P., \& Turok, N., 1994, Nature, 368, 27

\ref
Crittenden, R., Bond, J.R., Davis, R.L., Efstathiou, G., \& Steinhardt, P.,
1993, Phys. Rev. Lett., 71, 324

\ref
Danese, L., \& Partridge, R.B., 1989, ApJ, 342, 604

\ref
Davies, R.D., et al., 1992, MNRAS, 258, 605.

\ref De~Amici, G., et al. 1994,
in Proc. of the Workshop "The Cosmic Microwave Background", Capri (NA),
Italy, 20--24 September 1993. Astroph. Lett \& Comm., in press.

\ref
de Bernardis, P., et al. 1994, ApJ, 422, L33.

\ref
Dragovan, M., et al. 1994, ApJ, 427, L67.

\ref
Fisher, M.L., et al. 1992, ApJ, 388, 242.

\ref
Franceschini, A., Danese, L., De Zotti, G., \& Toffolatti, L. 1988, MNRAS,
233, 157

\ref
Franceschini, A., Toffolatti, L., Danese, L., \& De Zotti, G. 1989, ApJ,
344, 35

\ref
Franceschini, A., Mazzei, P., Danese, L., \& De Zotti, G. 1994, ApJ,
427, 140

\ref
Gaier, T., Schuster, J., Gundersen, J., Koch, T., Seiffert, M., Meinhold, P.,
\& Lubin, P., 1992, ApJ, 398, L1

\ref
Ganga, K., Cheng, E., Meyer, S., \& Page, L., 1993, ApJ, 410, L57

\ref
Gautier III, T.N., Boulanger, F., Perault, M., \& Puget, J.L. 1992,
AJ, 103, 1313

\ref
Gundersen, J.O., et al. 1993, ApJ, 413, L1.

\ref
Hancock, S., et al. 1994, Nature, 367, 333.

\ref
Hara, T., M\"ah\"onen, P., \& Miyoshi, S., 1993, ApJ, 414, 421

\ref
Haslam, C.G.T., Quigley, M.J.S., \& Salter, C.J. 1970, MNRAS, 147, 405

\ref
Haslam, C.G.T., et al. 1982, A\&AS, 47, 1

\ref
Hauser, M.G., 1993, Proc. of the Workshop "Back to the Galaxy",
College Park, Maryland, USA, October 1992.

\ref
Kamionkowsky, M., Spergel, D.N., \& Sugiyama, N. 1994, ApJ, 426, L57.

\ref
Kogut, A., et al. 1992, ApJ, 401, 1.

\ref
Lasenby, A.N., Davies, R.D., Hancock, S., Gutierrez de la Cruz, C.M., Rebolo
R., \& Watson, R.A. 1994, Proc. of the Workshop "Present and Future of
the Cosmic Microwave Background", Santander, Spain, 1993, eds. J.L. Sanz,
E. Martinez--Gonzalez \& L. Cayon, Springer-Verlag, p.91.

\ref
Lockman, F.J., Jahoda, K., \& McCammon, D., 1986, ApJ, 302, 432

\ref
Mandolesi, N., Smoot, G.F., Bersanelli, M., Cesarsky, C., Lachieze-Rey, M.,
Danese, L., Vittorio, N., De~Bernardis, P., Dall'Oglio, G., Sironi, G.,
Crane, P., Janssen, M., Partridge, B., Beckman, J., Rebolo, R.,
Puget, J.L., Bussoletti, E., Raffelt, G., Davies, R., Encrenaz, P.,
Natale, V., Tofani, G., Merluzzi, P., Toffolatti, L., Scaramella, R.,
Martinez-Gonzalez, E., Saez, D., Lasenby, A., \& Efstathiou, G. 1994
in Proc. of the Workshop "The Cosmic Microwave Background", Capri (NA)",
20--24 September 1993. Astroph. Lett \& Comm., in press.

\ref
Markevitch, M., Blumenthal, G.R., Forman, W., Jones, C., \& Sunyaev, R.A.
1992, ApJ, 395, 326

\ref
Mather, J.C., et al. 1994, ApJ, 420, 439

\ref
Meinhold, P.R., \& Lubin, P.M., 1991, ApJ, 370, L11

\ref
Meinhold, P.R., Chingcuanco, A.O., Gundersen, J.O., Schuster, J.A., Seiffert,
M.D., Lubin, P.M., Morris, D., \& Villela, T., 1993, ApJ, 406, 12

\ref
Meyer, S.S., Cheng, E.L., \& Page L.A., 1991, ApJ, 371, L7

\ref
Pan, S.K., et al. 1989, IEEE Trans. Microwave Theory Tech., 37, 3.

\ref
Perivolaropoulos, 1994, in Proc. of the Workshop "The Cosmic Microwave
Background", Capri (NA), 20-24 September 1993. Astrophys. Lett. \& Comm.,
in press

\ref
Pospieszalski, M.W., et al. 1993, IEEE MTT-S Digest, 515.

\ref
Pospieszalski, M.W., 1993, 23rd EuMC, p73.

\ref
Readhead, A.C.S., \& Lawrence, C.R., 1992, ARA\&A, 30, 653

\ref
Reich, W., 1982, A\&AS, 49, 219

\ref
Reich, P. \& Reich, W. 1986, A\&AS, 63, 205

\ref
Reich, P. \& Reich, W. 1988, A\&AS, 74, 7

\ref
Reynolds, R.J. 1992, ApJ, 392, L35

\ref
Rowan--Robinson, M. 1992, MNRAS, 258, 787

\ref
Schuster, J., et al. 1993, ApJ, 412, L47.

\ref
Scott, D., Srednicki, M., \& White, M., 1994, ApJ, 421, L5

\ref
Smoot, G.F., et al, 1992, ApJ, 396, L1.

\ref
Stark, A.A., et al., 1992, ApJS, 79, 77

\ref
Sugiyama, N., Silk, J., \& Vittorio, N., 1993, ApJ, 419, L1

\ref
Toffolatti, L., Danese, L., Franceschini, A., Mandolesi, N., Smoot, G.F.,
Bersanelli, M., Vittorio, N., Lasenby, A., Partridge, R.B., Davies, R.,
Sironi, G., Cesarsky, C., Lachieze--Rey, M., Martinez--Gonzalez, E.,
Beckman, J., Rebolo, R., Saez, D., de Bernardis, P., Dall'Oglio, G.,
Crane, P., Janssen, M., Puget, J.L., Bussoletti, E., Raffelt, G., Encrenaz,
P., Natale, V., Tofani, G., Merluzzi P., Scaramella, R., and Efstathiou, G.,
1994, in Proc. of the Workshop "The Cosmic Microwave
Background", Capri (NA), 20-24 September 1993. Astrophys. Lett. \& Comm.,
in press

\ref
Turok, N., \& Spergel, D.N., 1990, Phys. Rev. Lett., 64, 2736

\ref
Ue--Li Peh, \& Spergel, D.N., 1993, in Proc. of the Workshop
"The Cosmic Microwave Background", Capri (NA), 20-24 September 1993.
Astrophys. Lett. \& Comm., in press

\ref
White, M., Krauss, L.M., \& Silk, J., 1993, ApJ, 418, 535

\ref
Wright, E.L., et al., 1991, ApJ, 381, 200

\ref
Wright, E.L., et al., 1994, ApJ, 420, 450

\vfill\eject

\oneskip \noindent
\centerline{\bf FIGURE CAPTIONS}

\oneskip
\noindent
{\bf Figure 1.}  Estimated fluctuation levels of galactic emission
components at $\sim$ 0.5 degree angular
scale in high galactic latitude sky areas, $\vert b\vert\geq 40$
(in terms of the thermodynamic temperature $\Delta T/T$).
The horyzonthal continuous line
indicates a constant level $\Delta T/T= 10^{-5}$.
The dotted line represents the fluctuations due to the synchrotron
emission in regions where the rms intensity at 100 GHz is less than
$2.1\times 10^{-4}$ MJy/sr (see text).
As for the free--free emission, the short--dashed line refers to a $\Delta T/T$
level estimated assuming an rms intensity fluctuations of about 30$\%$
of the total emission (see \S 2.2.1).
The continuous line refers to fluctuations generated by the
galactic dust based on the assumption that
the rms variations amount to 100$\%$  of
the emission in the {\it coldest} areas of the sky, i.e. 0.5 MJy/sr
at 100 $\mu$m (see text);  we have also reported the contribution to the total
$\Delta T/T$ level given by the two
dust components (``cold'': dot--long--dashed; ``warm'': dot--short--dashed)
as modelled by Wright et al. (1991) (see \S 2.2 for more details).

\oneskip
\noindent
{\bf Figure 2.} {Estimated fluctuation levels due to extragalactic point
sources (in terms of the thermodynamic
temperature $\Delta T/T$). The plotted
curves refer to a source detection limit $x_c=5\sigma$ and
to different models for the evolution of the cold dust.
The higher $\Delta T/T$ level corresponds to the model by Franceschini et al.
(1988) assuming strong cosmological evolution of the most luminous
far--IR selected sources, while the lower one refers
to a moderate cosmological evolution of both late-- and early--type galaxies
(Franceschini et al., 1994).
Both models give integrated intensities $I(\nu)$ still compatible with
the recent $COBE$ FIRAS upper limits on the CBR residuals in the sub--mm domain
(Mather et al., 1993; Wright et al. 1993) but the higher one is close to
infringe the FIRAS limits. The frequency channels foreseen for the COBRAS/SAMBA
experiment, 31.5, 53, 90, 125, 230, 375 and 670 GHz, are also indicated
by the dotted vertical lines.}

\oneskip
\noindent
{\bf Figure 3.} {Comparison of the spectral coverage of some
anisotropy experiments recently performed or under study.
Dark squares represent bands covered with radiometric techniques,
while boxes are bolometric passbands. For details on the single
experiments see, e.g., UCSB-SP91: Schuster et al. 1994;
Python: Dragovan et al. 1994; Tenerife: Hancock et al. 1994;
MAX: Gundersen et al. 1993; ARGO: De Bernardis et al. 1994;
TopHat: Cheng 1994; $COBE$-DMR: Smoot et al. 1992;
COBRAS/SAMBA: Mandolesi et al. 1994.
The Tenerife and $COBE$-DMR measurements have been performed at
$\sim 5^{\circ}$ and $7^{\circ}$ angular resolution whereas
all the other experiments are at degree or sub--degree angular scale.

\vfill\eject

\def\binit{\hoffset=-.3 truecm
	    \voffset=-1. truecm
	    \hsize=16.5 truecm
	    \vsize=25.5 truecm
	    \baselineskip=8.pt
	    \lineskip=0pt
	    \lineskiplimit=0pt}

\nopagenumbers
\magnification=\magstep1
\binit
\bigskip\bigskip
\centerline{\bf TABLE I. Space vs. Ground and Balloon Experiments}
\bigskip
\hrule height1pt\hsize=16.5 truecm \vskip2pt \hrule\hsize=16.5 truecm
\medskip
{\sl SCIENTIFIC OBJECTIVES: extensive imaging (i.e., more than $\sim 40\%$)
of the sky (in particular the portion at $\vert b\vert >40^o$)
with a sensitivity $\Delta T/T \simeq 3\times 10^{-6}$ per pixel
and angular resolution $\sim$10$^{\prime}$-30$^{\prime}$.}
\medskip
\hrule height1pt\hsize=16.5 truecm \vskip2pt \hrule\hsize=16.5 truecm
\smallskip\smallskip
\settabs 4 \columns
{\bf \+\hfill &GROUND\hfill &BALLOON\hfill& SPACE\hfill \cr}
\smallskip\smallskip
\hrule\hsize=16.5 truecm
\smallskip\smallskip
\+FREQUENCY \hfill & \hfill & \hfill & \hfill \cr
\smallskip\smallskip
\+freq. range\hfill &10-18,28-40,90 GHz\hfill &10-900 GHz\hfill &10-900
GHz\hfill \cr
\smallskip\smallskip
\+technique\hfill &radiometers\hfill &bolometers\hfill&both\hfill \cr
\+\hfill &\hfill &and/or radiometers\hfill&\hfill \cr
\smallskip\smallskip
\hrule\hsize=16.5 truecm
\smallskip\smallskip
\+ENVIRONMENT\hfill &\hfill &\hfill &\hfill \cr
\smallskip
\+thermal variability\hfill &several K/day\hfill &few K/day\hfill
&$\sim 0.1$ K/few weeks (L5)\hfill \cr
\smallskip\smallskip
\+diffraction, reflection \hfill &moving parts\hfill &balloon, gondola,\hfill
&payload \hfill \cr
\+emission \hfill &on ground\hfill &chopping-spinning \hfill &spacecraft\hfill
\cr
\smallskip\smallskip
\+required sidelobe\hfill& \hfill & \hfill &\hfill \cr
\+rejection (Earth)\hfill& $10^{13}$\hfill & $10^{13}$\hfill
&$10^{9}$ (L5)\hfill \cr
\smallskip\smallskip
\hrule\hsize=16.5 truecm
\smallskip\smallskip
\+DATA\hfill &\hfill &\hfill &\hfill \cr
\smallskip
\+$T_{\rm {atm}}/\Delta T_{\rm {CMB}}$\hfill &$\geq 10^6$\hfill
&$\sim 10^3$\hfill &0\hfill \cr
\smallskip\smallskip
\+Data rejected for\hfill &\hfill &\hfill &\hfill \cr
\+atmospheric \hfill &$\sim$ 70\%\hfill &$\sim$ 10\% \hfill &None \hfill \cr
\+fluctuations\hfill &\hfill &\hfill &\hfill \cr
\smallskip\smallskip
\+actual integration &$\sim$ 10-20\%\hfill &$\sim$ 3-4\%\hfill &$\geq
90$\%\hfill \cr
\+time (per year)\hfill &\hfill &\hfill &\hfill \cr
\smallskip\smallskip
\+overall efficiency\hfill &low\hfill &medium\hfill &very high\hfill \cr
\smallskip\smallskip
\hrule\hsize=16.5 truecm
\smallskip\smallskip
\+FOREGROUNDS\hfill &\hfill &\hfill & \hfill \cr
\smallskip\smallskip
\+Galactic\hfill &hampered by \hfill &reliable only with \hfill
&highly reliable\cr
\+foreground \hfill &atmospheric fluct.,\hfill &many frequencies\hfill
& (multifreq. obs. + \cr
\+subtraction \hfill & limited sky coverage,\hfill &and many flights\hfill &
sky coverage) \hfill\cr
\+\hfill &small freq. range\hfill &\hfill & \hfill \cr
\smallskip\smallskip
\+extragalactic\hfill &external\hfill &external\hfill & multifreq. obs. +\hfill
\cr
\+sources\hfill &surveys\hfill &surveys\hfill & external surveys\hfill \cr
\smallskip\smallskip
\hrule\hsize=16.5 truecm
\smallskip\smallskip
\+SKY COVERAGE\hfill&\hfill &\hfill &\hfill \cr
\smallskip
\+ pixels per year at\hfill&\hfill &\hfill &\hfill \cr
\+$\Delta T/T\sim 3\times 10^{-6}$\hfill & $\sim$ 50-200 pixels\hfill &
$\sim 1000$ pixels\hfill & $\sim 100,000$ pixels\hfill \cr
\+(per frequency)\hfill & \hfill & \hfill & \cr
\smallskip\smallskip
\+\% of all sky\hfill & 0.1\%\hfill & 0.6\%\hfill & $\simeq$ 60\%\cr
\smallskip\smallskip
\+limiting factors\hfill &data efficiency,\hfill &integration time,\hfill
&instrument sensitivity,\hfill \cr
\+\hfill &accessibility, \hfill &accessibility, \hfill &mission
lifetime\hfill \cr
\+\hfill &foreground subtrac.\hfill &foreground subtrac.\hfill &\hfill \cr
\smallskip\smallskip
\hrule \hsize=16.5 truecm \vskip2pt \hrule height1pt\hsize=16.5 truecm
\smallskip\smallskip

\bye